 \documentclass{aa}  
\usepackage{graphicx}
\usepackage{txfonts}
\usepackage{natbib}
\begin{document} 

\title{Statistical study of uncertainties in the diffusion rate of species on interstellar ice and its impact on chemical model predictions}
%
%
%
%
\author{Wasim Iqbal 
        \and Valentine Wakelam 
        \and Pierre Gratier  
        }
\institute{Laboratoire d'astrophysique de Bordeaux, Univ. Bordeaux, CNRS, B18N, all\'ee Geoffroy Saint-Hilaire, 33615 Pessac, France\\
          \email{wasimiqbal2009@gmail.com}\\
          \email{valentine.wakelam@u-bordeaux.fr}\\
          \email{pierre.gratier@u-bordeaux.fr }
           }
\date{Received ------------; accepted -------}
%
%
\abstract
 {Diffusion of species on the dust surface is a key process for determining the chemical composition of interstellar ices. On the dust surface, adsorbed species diffuse from one potential well to another and react with other adsorbed reactants, resulting in the formation of simple and complex molecules.}
%
 {We study the impact on the abundances of the species simulated by the chemical codes by considering the uncertainties in the diffusion energy of adsorbed species. We aim to limit the uncertainties in the abundances as calculated by chemical codes by identifying the surface species that result in a larger error because of the uncertainties in their diffusion energy.}
%
%
 {We ran various cases with 2000 to 10000 simulations in each case and varied the diffusion energies of some or all surface species randomly. We calculated Pearson correlation coefficients between the abundances and the ratio of diffusion to binding energy of adsorbed species. We identified the species that introduce maximum uncertainty in the ice and gas-phase abundances. With these species we ran three sets, with 2000 simulations in each, to quantify the uncertainties they introduce.}
%
%
 {We present the abundances of various molecules in the gas phase and also on the dust surface at different time intervals during the simulation. We show which species produce a large uncertainty in the abundances. We sorted species into different groups in accordance with their importance in propagating uncertainty in the chemical network.}
%
%
 {We show that CO, H$_2$, O, N, and CH$_3$ are the key species for uncertainties in the abundances, while CH$_2$, HCO, S and O$_2$ come next, followed by NO, HS, and CH. We also show that by limiting the uncertainties in the ratio of diffusion to binding energy of these species, we can eliminate the uncertainties in the gas-phase abundances of almost all the species.}
\keywords{astrochemistry -- ISM: clouds -- ISM: molecules -- ISM: abundances -- molecular processes -- diffusion energy}
\titlerunning{Diffusion rate and uncertainties in the chemical abundances}
\authorrunning{Iqbal et al.}
\maketitle
\section{Introduction}
\label{intro}
The role of interstellar dust grains in the formation of simple and complex molecules in the interstellar medium (ISM) is of fundamental importance \citep{Gould1963, Stecher1966, Hollenbach1970, Williams1971, Watson1972, Watson1976, Hasegawa1992}. Therefore, to better understand the observed abundances in the different regions of the ISM, astrophysical models such as the Nautilus gas-grain code \citep{Ruaud2016} have been developed to simulate the chemistry on the grain surfaces coupled with the gas-phase chemistry. On the grain surface, there are two types of surface reaction mechanisms: the Langmuir-Hinshelwood (LH) mechanism, and the Eley-Rideal (ER) mechanism. In the LH mechanism, reactions occurs between two physisorbed species. Reactions takes place when two reactive species encounter each other through the process of diffusion, provided there is no barrier for reaction. The reaction is only possible if at least one of the reactive species is able to overcome the diffusion barrier and thus is able to diffuse and find the other reactive species. In the ER mechanism, reaction takes place when reactive gaseous species directly collide with an adsorbed reactive species, provided there is no barrier for the reaction. In most models, the LH mechanism is considered to be the dominating mechanism as the ER mechanism is found to be inefficient because the density of reactive species on the surface of the grain is lower \citep{Ruaud2016}. 

Since the LH mechanism is a diffusive mechanism, it is obvious that the diffusion of the adsorbed species determines the possible chemical pathways on the grain surface. Thus diffusion strongly influences the ice composition. Despite its importance, diffusion of the adsorbed species (in the astrophysical context) is a poorly understood process. In the literature, some studies, both experimental and theoretical, have been made to estimate the diffusion rate of some species \citep{Livingston2002, Al-Halabi2007, Watanabe2010, Mispelaer2013, Karssemeijer2014}. \citet{Livingston2002} measured the bulk diffusion and the surface diffusion for a number of molecular species for a temperature range between 140K and 200K. In their experiments, they used crystalline ice as the diffusion medium. Extrapolation of their results to 10K (dense cloud conditions) gives undesired values and thus is not very useable in the astrophysical context. \cite{Al-Halabi2007} and \cite{Watanabe2010} studied the diffusion of H atoms on the amorphous solid water ice. \cite{Mispelaer2013} performed experimental measurements for diffusion rates of CO, HNCO, H$_2$CO, and NH$_3$ in amorphous water ice between a temperature range of 35K and 140K with 90\% uncertainty on each diffusion coefficient. Because of the large correlated uncertainties in the parameters, they claimed that it is not possible to extrapolate their result with confidence to lower temperatures, which would be more relevant to astrochemistry. \cite{Karssemeijer2014} reported computational calculations for the diffusion-desorption ratio of adsorbed CO and CO$_2$ on water ices. These studies showed that the diffusion rate varies with the adsorbed species and strongly depends on the binding surface and the surface temperature. The diffusion rates of a species are also different in the bulk and on the surface. Current chemical models generally have more than 200 surface species. The number of the surface species and the surface reactions indeed will only increase with time. The poor understanding of the diffusion energy of most of the surface species in the astrochemical models seriously limits the confidence in the accuracy of results obtained with these models.

 In chemical models, the diffusion energy ($E_\text{d}(i)$)  for any species $i$ is often taken to be a fraction of its binding energy ($E_\text{b}(i)$) to the surface, and for simplicity, the ratio $E_\text{d}(i)$/$E_\text{b}(i)$ is taken to be the same for all the surface species. The reason is again that the diffusion of surface species is still a poorly understood process. In the literature we find many values of this ratio, ranging from 0.3 to 0.8 \citep{Watson1972, Watson1976, Hasegawa1992, Biham2001, Chang2005, Ruaud2016}. In the absence of a proper understanding of the diffusive process, however, it is not possible to treat the diffusion energy of each species independently. It is therefore beyond doubt that the maximum possible error or the uncertainty in the calculation of the diffusion rate of any species on the grain surface comes, for a large part, from the uncertainty in the assumed value of the $E_\text{d}(i)$, the height of the diffusion barrier. 

 In this work we try to explore the order of uncertainties and its impact on the simulated abundances of species that are observed in dark dense clouds such as TMC1.
The paper is organized as follows. In Sect.~\ref{sec:model} we describe our approach to the problem in detail and also present the Nautilus gas-grain model and the chemical network used in our simulations. In Sect.~\ref{results} we present our results obtained using the Nautilus gas-grain model with the artificially but statistically generated surface parameters.  In this section, we compute and compare the uncertainties in the abundances of species in different cases. In Sect. \ref{R_sec:5} we compare our model with observations in the TMC-1 and L134N clouds. We also compare these results with the standard Nautilus model. We offer our final conclusions in the last section. 

\section{Chemical modeling and method}
\label{sec:model}
\subsection{Chemical network and the simulation model}
\label{sec:nautilus}
In our simulations, we used the gas-phase chemistry, which is based on the kida.uva.2014 public network \citep{Wakelam2015}. We used the same surface chemistry as in \citet {Ruaud2016}. To simulate the chemical network, we used the Nautilus gas-grain code, which is based on the rate equation approximation (\citet{Hasegawa1992, Hasegawa1993b}). We used the three-phase version of Nautilus \citep[see][]{Ruaud2016}. In the three-phase model we have the gas phase, the grain surface, and the grain mantle, such that the gas phase is coupled with the grain surface and the grain surface is coupled with the grain mantle. Exchange of species is possible between either the gas phase and the grain surface or the grain surface and the grain mantle. There is no direct interaction between the grain mantle and the gas phase. In the model, we have chemistry between different species in all the three phases. On the grain we have the physisorption of neutral species on the surface, the diffusion of these species, and their reactions and thermal desorption. In the mantle we have diffusion and reaction, but thermal desorption of species is not possible. The surface is defined as the top two monolayers of species. Thus in the case of desorption from the surface, species from the mantle come on the top and form the new surface layer, and similarly, new mantle layers are formed from the surface species through accretion of new species on the surface. In addition to the accretion, diffusion, recombination, and thermal desorption, we also consider nonthermal desorption processes such as cosmic-ray-induced desorption, UV (direct and indirect) photodesorption, and chemical desorption. Details of all the processes included in our chemical model can be found in \citet{Ruaud2016}. 
\subsection{Grain surface reaction mechanism}
\label{sec:surf_reaction}
In our model we consider only the LH mechanism. The surface reaction rate between species $i$ and $j$ for the LH mechanism is given by
\begin{equation}
 k^s_{ij} = \kappa^s_{ij}\Bigg(\frac{1}{t^s_\text{hop}(i)} + \frac{1}{t^s_\text{hop}(j)}\Bigg)\frac{1}{N_\text{site}n_\text{dust}}~ [\text{cm}^3\text{s}^{-1}],
 \label{eqn:1}
 \end{equation}
where the superscript $s$ represents the surface reaction, $\kappa_{ij}$ is the probability of reaction \citep{Chang2007}, $N_\text{site}$ is the number of binding sites on the grain surface, $n_\text{dust}$ is the number density of dust grains, and $t_\text{hop}(i)$  is the thermal hopping time of species $i$.

Following \cite{Hasegawa1992}, we used $\kappa^s_{ij} = 1$ if the reaction is exothermic and barrierless. For exothermic reactions with activation barriers ($E_A(i,j)$),  however, we calculated $\kappa^s_{ij}$, following the method described in \cite{Chang2007}, considering the competition among reaction, hopping, and evaporation. In this case, $\kappa^s_{ij}$ is given as
\begin{equation}
\kappa^s_{ij} = \frac{\nu^s_L p^s_\text{tun}(i,j)}{\nu^s_L p^s_\text{tun}(i,j) + k^s_\text{hop}(i) + k^s_\text{evap}(i) + k^s_\text{hop}(j) + k^s_\text{evap}(j)},\end{equation}
where $p^s_\text{tun}(i,j)$ is the quantum-mechanical probability for tunneling through a rectangular barrier of thickness $a$ ($a$ is taken to be 1 \AA), $k^s_\text{hop}(i)$ and $k^s_\text{evap}(i)$ are hopping and evaporation rates for species $i$. $\nu^s_L$ is the higher value among $\nu^s_i$ and $\nu^s_j$  \citep{Garrod2011}, where $\nu^s_i$ is called the characteristic vibration frequency of the species $i$ \citep[see][]{Hasegawa1992}.  $p^s_\text{tun}(i,j)$ is calculated as  
\begin{equation}
p^s_\text{tun}(i,j)=\exp[-2a/ \hbar\sqrt{2\mu E_A(i,j)}],
\end{equation}
where $\mu$ is the reduced mass \citep[see][for details]{Hasegawa1992}.

The hopping and evaporation rates of a species $i$ are given by
 \begin{equation}
  k^x_\text{hop}(i)=  \frac{1}{t^x_\text{hop}(i)} = \nu^x_i \exp\Bigg(\frac{-E_\text{d}^x(i)}{T_\text{dust}}\Bigg), 
 \end{equation}
 and
 \begin{equation}
  k_\text{evap}(i)=  \frac{1}{t_\text{evap}(i)} = \nu^s_i \exp\Bigg(\frac{-E_\text{b}^s(i)}{T_\text{dust}}\Bigg), 
 \end{equation}
 respectively, 
  where $T_\text{dust}$ is the dust temperature, and the superscript $x = (s$ or $m)$ denotes the surface or the mantle as the diffusion energy is different for the surface and the bulk species. 
 
 To calculate the reaction rate ($k^m_{ij}$) in the grain mantle, $k^s_{ij}$ in Eqn.\ref{eqn:1} is divided by $\sum_i N_m(i)/N_\text{site}$ , where $N_m(i)$ is the total  number of species $i$ in the mantle \citep[see][]{Ruaud2016}. 

In the standard Nautilus model we have 225 surface species, and for all species, the ratio $E_\text{d}(i)/E_\text{b}(i)$ (we call this ratio $\mu^x$, where again $x= (s$ or $m)$ represents the surface and the mantle, respectively) is kept constant at 0.4 on the surface, and in the mantle it is 0.8. We made the fundamental assumption that the value of $\mu^s$ is different for all species and that its value lies between 0.25 to 0.75. We kept the possible minimum value of $\mu^s$ at 0.25 as some species can have very high mobility on certain ices \citep{Watanabe2010}. However, in the bulk, the mobility of the species is assumed to be much lower than on the surface. We therefore must have a high value of $\mu^m$ . This means that the window of uncertainty on the value of $\mu^m$ is smaller. For simplicity we kept the value of $\mu^m$ constant at 0.8 for all the species. 

 Given that we have 225 surface species and all species can have any value of $\mu^s$ between 0.25 to 0.75, it is not possible to simulate all possible values of $\mu^s$ for all surface species. We therefore ran 10000 simulations, and in each simulation, we randomly assigned values of $\mu^s$ for all species except of H. For H we kept the diffusion energy constant at 230K ($\mu^s = 0.35$) following the works of \cite{Al-Halabi2007} and \cite{Watanabe2010} on the diffusion of H atoms on the amorphous solid water ice.
\subsection{Other model parameters and binding energies}
\begin{table}
   \begin{center}
   \caption{Some important parameters used in our models.}
   \label{tbl:cold_dense_cloud_model}
   \begin{tabular}{@{}lll}
   \hline
   \hline
   Parameters                     & Value\\
   \hline
   $T_{gas}$                      & 10 K \\
   $n_\textrm{H}$                 & $2\times10^4$ cm$^{-3}$ \\
   $A_V$                          & 15 \\
   Cosmic ray ionization rate    & $1.3\times10^{-17}$ s$^{-1}$ \\
   Grain surface site density     & $8.0\times10^{14}$  cm$^{-2}$   \\
   \hline
   \end{tabular}
   \end{center}  
\end{table}
\begin{table}
   \begin{center}
   \caption{Elemental abundances and initial abundances.}
   \label{tbl:initial-abundances}
   \begin{tabular}{@{}lll}
   \hline
   \hline
   Element                      &       Abundance relative to H    & References  \\
   \hline
   H$_2$                        &       0.5                        &             \\
   He                           &       0.09                       &$^\textrm{a}$ \\
   N                            &       6.2$\times10^{-5}$         &$^\textrm{b}$ \\
   O                            &       2.4$\times10^{-4}$         &$^\textrm{c}$ \\      
   C$^+$                        &       1.7$\times10^{-4}$         &$^\textrm{b}$ \\   
   S$^+$                        &       8.0$\times10^{-9}$         &$^\textrm{d}$ \\      
   Si$^+$                       &       8.0$\times10^{-9}$         &$^\textrm{d}$ \\
   Fe$^+$                       &       3.0$\times10^{-9}$         &$^\textrm{d}$ \\
   Na$^+$                       &       2.0$\times10^{-9}$         &$^\textrm{d}$ \\
   Mg$^+$                       &       7.0$\times10^{-9}$         &$^\textrm{d}$ \\   
   P$^+$                        &       2.0$\times10^{-10}$        &$^\textrm{d}$ \\      
   Cl$^+$                       &       1.0$\times10^{-9}$         &$^\textrm{d}$ \\ 
   ice                          &       0                          &              \\
   \hline
   \end{tabular}
   \end{center} 
   $^\textrm{a}$\citet{Wakelam08}, $^\textrm{b}$\citet{Jenkins09}, 
   $^\textrm{c}$\citet{Hincelin2011}, $^\textrm{d}$\citet{Graedel82}   
\end{table}
We used parameters that are suitable for cold-core conditions (see Table~\ref{tbl:cold_dense_cloud_model}). These parameters were kept the constant in all simulations. In Table \ref{tbl:initial-abundances} we list the initial abundances we used in our all simulations. At the start of each simulation, we kept the ice abundance to zero.

We used binding energies of all species from the KIDA database. Most of the binding energies are from the original OSU database and are listed in KIDA (which can be found at http://kida.obs.u-bordeaux1.fr with the reference of the OSU database). For some species we used new binding energies from \cite{Ruaud2015}. Binding energies of H and H$_2$ were taken from \cite{Wakelam2017}, and that of N and O were from \cite{Tielens1982} and \cite{Tielens1987}, respectively. These binding energies are also available at http://kida.obs.u-bordeaux1.fr. with the respective references.
\section{Results}
\label{results}
\subsection{Uncertainty in $\mu^s$ and its effect on abundances}
\label{R_sec:1}
\begin{figure}
    \centering
   \includegraphics[width=0.5\textwidth,trim = 0cm 0cm 0cm 0cm, clip,angle=0]{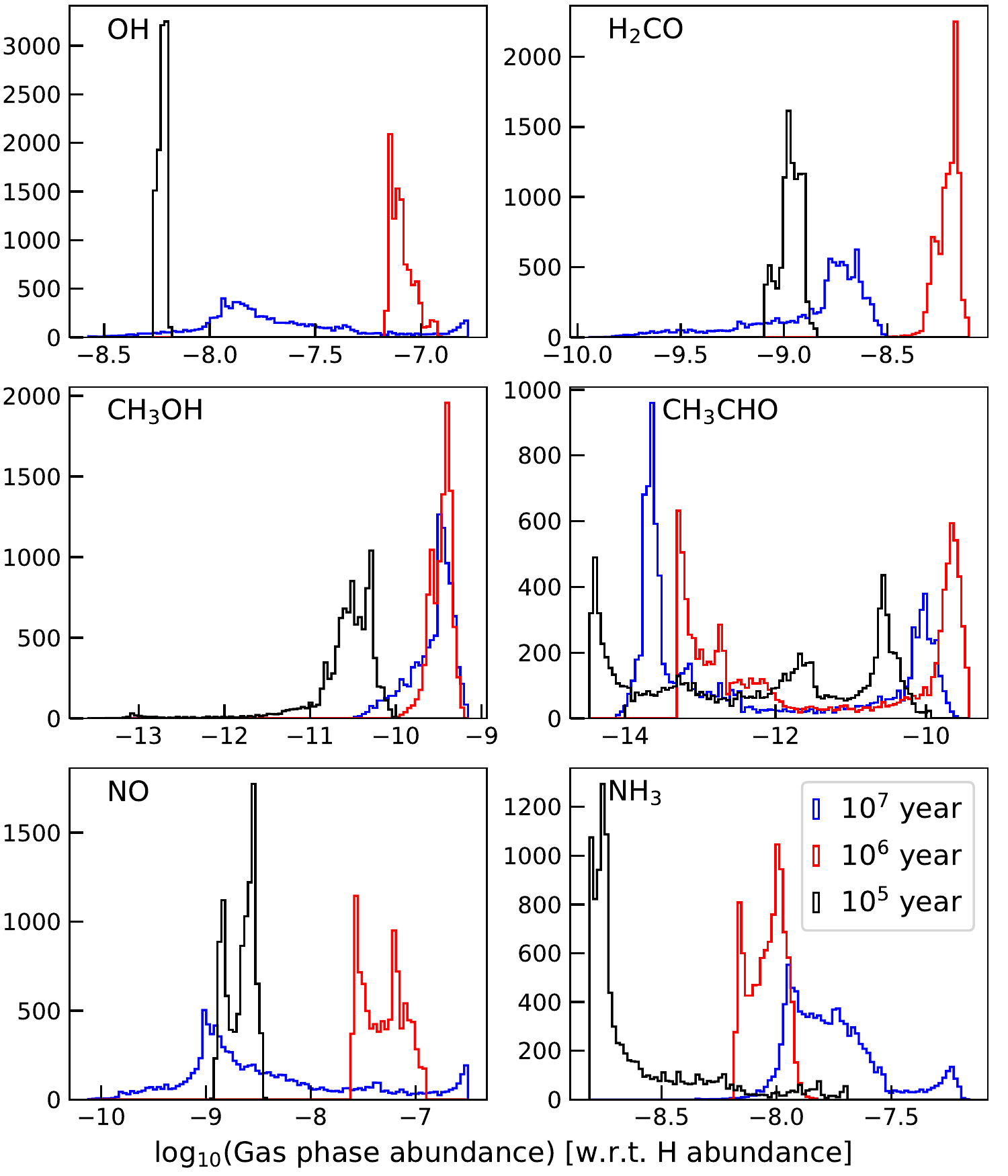}
   \caption{Histogram or distribution of abundances as obtained in 10000 simulations at three different times.}
   \label{Fig:1a}
\end{figure}
\begin{figure*}
    \centering
   \includegraphics[width=1\textwidth,trim = 0cm 4.4cm 0cm 4.3cm, clip,angle=0]{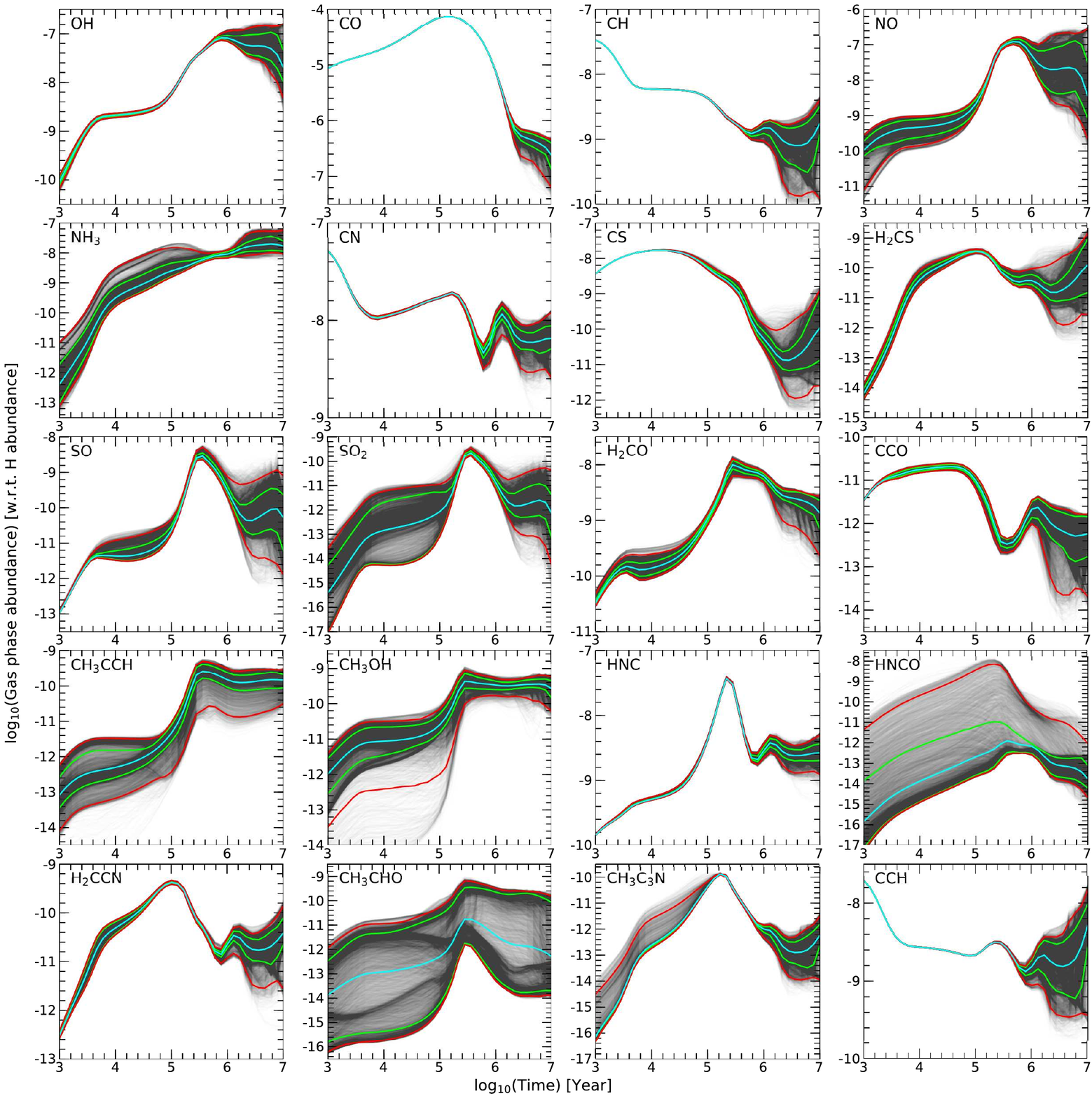}
   \caption{Gas-phase abundances of selected species as a function of time. Red lines show the 2 $\sigma$ deviation, green lines show 1 $\sigma$ deviation, and the cyan line shows the mean abundance. The intensity of black within the plot shows the abundance density distribution in 10000 simulations.}
   \label{Fig:1}
\end{figure*}
\begin{figure*}
    \centering
   \includegraphics[width=1\textwidth,trim = 0cm 4.4cm 0cm 4.3cm, clip,angle=0]{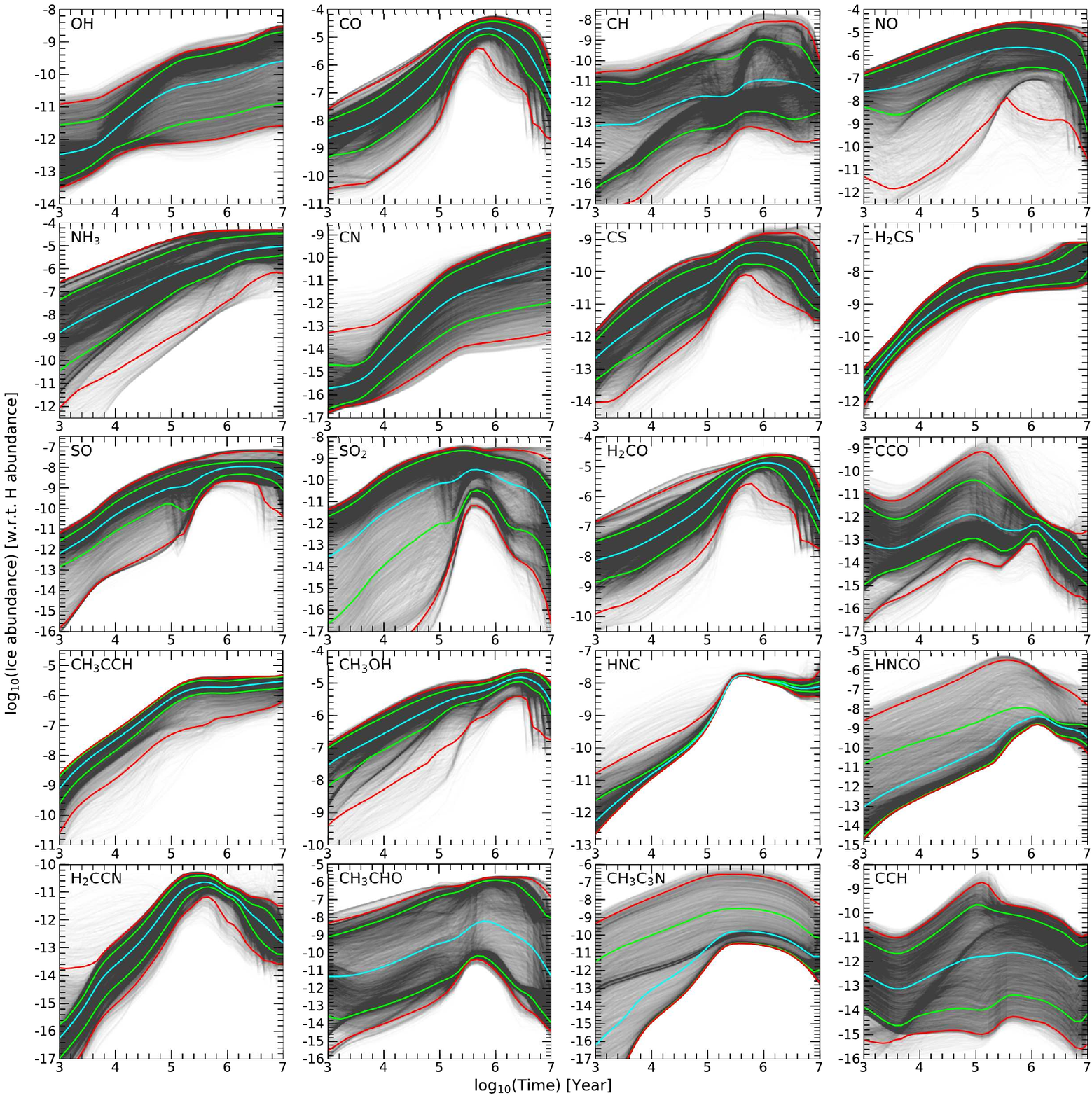}
   \caption{Total ice abundances of selected species on dust grains as a function of time. Red lines show the 2 $\sigma$ deviation, green lines show 1 $\sigma$ deviation, and the cyan line shows the mean abundance. The intensity of black within the plot shows the abundance density distribution in 10000 simulations.}
   \label{Fig:2}
\end{figure*}
We first consider the uncertainties on the species abundances that are due to the uncertainty in $\mu^s$ of all surface species. We ran 10000 simulations, and in each simulation, we randomly generated independent values of $\mu^s$ (such that $0.25 \leq \mu^s \leq 0.75$) for each species except for H, with a normal distribution centered on 0.5 and with a standard deviation of 1 $\sigma$. This uncertainty strongly depends on the time. We then considered times between $10^3$ and $10^7$ years.

In Fig. \ref{Fig:1a} we show the abundance histograms for selected species at three time intervals for all 10000 simulations. The distribution of the computed abundances (in log$_{10}$) is often peaked but not symmetrical. The distribution asymmetry is of various nature and strongly depends on time. Some species such as OH, H$_2$CO and NO show a narrow profile (a $2 \sigma$ variation of less than 0.5 (in log$_{10}$) in the abundance) in the distributions even up to $10^6$ years. This means that the uncertainties on their abundances is very small. After $10^6$ years, however, the uncertainties increase rapidly and the spread becomes more than two orders of magnitude. CH$_3$CHO shows large uncertainties in its distribution at all times. Furthermore, all distributions are asymmetric in nature, and some have more than one peak. For species such as CH$_3$OH and H$_2$CO, these multiple peaks are very close and it is possible to fit the distribution with a single Gaussian function, but for species such as NO, NH$_3$, or CH$_3$CHO, the peaks are well separated and must be fit with two or more Gaussian functions. For simplicity and also to use a single method for calculating the uncertainties in the abundances of all species, we used a percentile method to calculate the abundances within 1 $\sigma$ and 2 $\sigma$ deviation. We used the percentile function of numpy in python version 2.7 to calculate the 2.3, 15.9, 50, 84.1, and 97.7 percentile of the 10000 values of log$_{10}(X_i(t))$, where $X_i(t)$ is the abundance of the species $i$ at time $t$. All the abundances between 97.7 percentile and 2.3 percentile give us a 2 $\sigma$ interval on the two sides of 50 percentile value. Similarly, the range between the 84.1 percentile and 15.9 percentile of abundances gives us a 1 $\sigma$ interval. We can also define an error $(\delta)$ in the abundances as half of the total uncertainties in the abundances. Thus we have
\begin{equation}
 \delta_{1\sigma}({\rm{}log}_{10}(X_i(t)))=\frac{1}{2}[{\rm{}log}_{10}(X_i(t)_{84.1p})-{\rm{}log}_{10}(X_i(t)_{15.9p})]
 \label{eqn:1sigma}
 \end{equation}
 \begin{equation}
 \delta_{2\sigma}({\rm{}log}_{10}(X_i(t)))=\frac{1}{2}[{\rm{}log}_{10}(X_i(t)_{97.7p})-{\rm{}log}_{10}(X_i(t)_{2.3p})]
 \label{eqn:2sigma}
 .\end{equation}
 We note that the asymmetry of the abundance profiles implies that the mean abundance ($\overline{X_i(t)} = \sum_i X_i(t)/N$, where $N$ is the number of simulations) computed from the simulations does not necessarily represent the favored value.

In Fig. \ref{Fig:1} we plot the gas-phase abundances of 20 species observed in the dark cloud TMC-1 (CP). We also show the 1 $\sigma$ (green lines) and 2 $\sigma$ (red lines) deviation and the mean abundance (cyan line) for each species. All the species in Fig. \ref{Fig:1} can be divided into two groups. The first group is composed of OH, CO, CH, CN, CS, HNC, and CCH. These species show a negligible variation in their gas-phase abundance until as late as $5\times 10^5$ years. After this, the variation in their abundance starts to increase rapidly. These species are essentially those that form in the gas phase, and thus the variations in their abundances in the ice may not correlate with their gas-phase abundances. To verify this, we show in Fig. \ref{Fig:2} that these species have indeed large variations in their ice abundances even at early time in the simulation, but the ice abundances are smaller than those of the gas-phase by more than one order. Desorption of these species from the surface therefore does not produce large variation in their gas-phase abundances.  
At later time, however, when the gas is depleted in heavier elements as a result of freeze-out on dust grains, the formation efficiency of these species in the gas phase is greatly reduced.  By this time, the ice abundances become significant, and thus the balance between the freeze-out and the non-thermal desorption processes controls the amount of these species in the gas phase, which causes the observed correlation between the variations in the ice abundances and the gas- hase abundances.

In the second group we collected all other species. These species show significantly large variations in their abundances from very early time in the simulation. This trend indicates that either these species are predominantly formed on the grain surface and the uncertainty in $\mu^s$ of surface species directly translates into the variation in their abundances, or that the gas-phase abundances of these species strongly depend on other species that predominantly form on the grain surface. One point to note here is that the uncertainty does not always propagate positively with time. This means that we can have a maximum and minimum uncertainty at any time in the simulation. As we clearly see in Fig. \ref{Fig:1}, SO$_2$, CH$_3$OH and CH$_3$CHO have the maximum uncertainties in abundances at around $5\times 10^4$ years and the minimum uncertainties at around $5\times 10^5$ years. This behavior strongly depends on whether the species is predominantly formed in the gas phase or on the grain surface, and in the simulation, this may change with time. For example, at the beginning of the simulation, the gas-phase SO$_2$ is predominantly formed through the grain surface reaction
\begin{equation*}
\rm JO + JSO \rightarrow SO_2, 
\end{equation*}
where the prefix J represents the gain surface reaction. By $5\times 10^5$ years, however, SO$_2$ is predominantly formed through the  gas phase reactions\begin{equation*}
\rm OH + SO \rightarrow H + SO_2 ,
\end{equation*}
\begin{equation*}
\rm HSO_2^+ + e^- \rightarrow H + SO_2
\end{equation*}
and
\begin{equation*}
\rm O + SO \rightarrow SO_2.
\end{equation*}
Similarly, at the beginning, CH$_3$OH is only formed on the grain surface through the reactions
\begin{equation*}
\rm JH + JCH_3O \rightarrow CH_3OH        
\end{equation*}
and
\begin{equation*}
\rm JH + JCH_2OH \rightarrow CH_3OH,        
\end{equation*}
but around $5\times 10^5$ years, it is also formed in the gas phase with significant efficiency thorugh the reactions
\begin{equation*}
\rm CH_3OCH_3 + C \rightarrow CH_3OH + C_2H_2,        
\end{equation*}
\begin{equation*}
\rm CH_3OH_2^+ + e^- \rightarrow H + CH_3OH,        
\end{equation*}
\begin{equation*}
\rm CH_3OCH_4^+ + e^- \rightarrow CH_3 + CH_3OH,        
\end{equation*}
and
\begin{equation*}
\rm CH_3OCH_3^+ + e^- \rightarrow CH_2 + CH_3OH,        
\end{equation*}
although it is still formed predominantly on the grain surface. We found a similar trend for other species as well. This is no surprise as the uncertainty in $\mu^s$ produces the uncertainty in the grain surface reaction rates alone. Therefore we expect a higher uncertainty when a certain species is predominantly formed on the grain surface, a reduced uncertainty when the species is formed significantly in the gas phase, and absolutely no uncertainty when it is solely formed in the gas phase.
%
%

\subsection{Identification of key species}
\label{R_sec:2}
\begin{figure}
    \centering
   \includegraphics[width=0.5\textwidth,trim = 0cm 0cm 0cm 0cm, clip,angle=0]{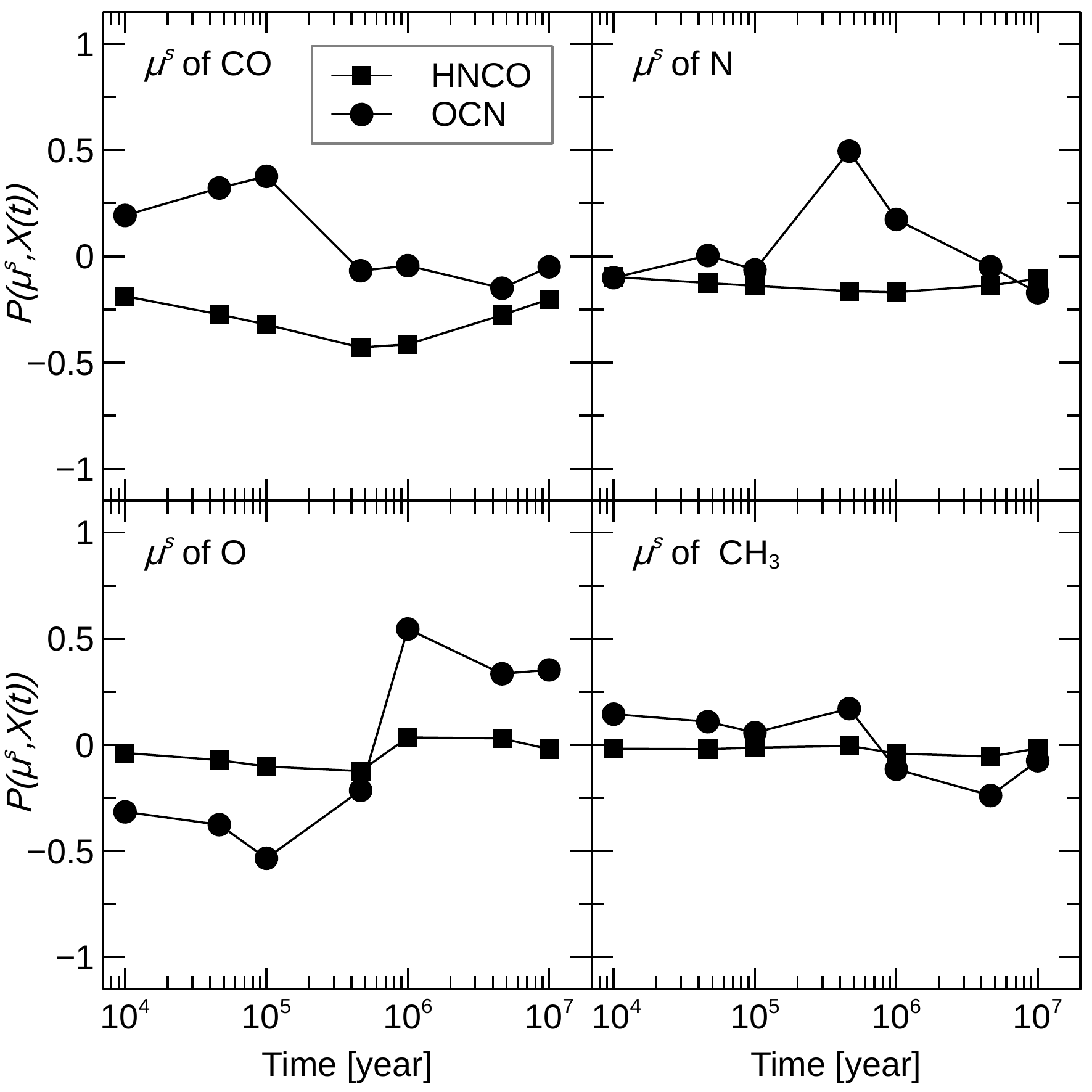}
   \caption{Calculated Pearson correlation coefficient (at different times) between the gas-phase abundances of HNCO and OCN and $\mu^s$ of CO, N, O, and CH$_3$}
   \label{Fig:3}
\end{figure}
\begin{table}
   \begin{center}
   \caption{Summary of the different cases.}
   \label{tbl:case}
   \begin{tabular}{@{}lll}
   \hline
   \hline
   Case     &       Species with constant $\mu^s$                        & Number of simulations  \\
   \hline
   Case A   &      Only H                                                &  10000           \\
   Case B   &      H, H$_2$, N, O, CO, and CH$_3$                         &  2000            \\
   Case C   &      All species in case B plus                            &  2000            \\
            &      CH$_2$, HCO, S, and O$_2$                              &                  \\
   Case D   &      All species in cases A and B                           &  2000            \\
            &      plus NO, HS, and CH                                    &                  \\
   \hline
   \end{tabular}
   \end{center} 
\end{table}
\begin{figure*}
    \centering
   \includegraphics[width=1\textwidth,trim = 0cm 0cm 0cm 0cm, clip,angle=0]{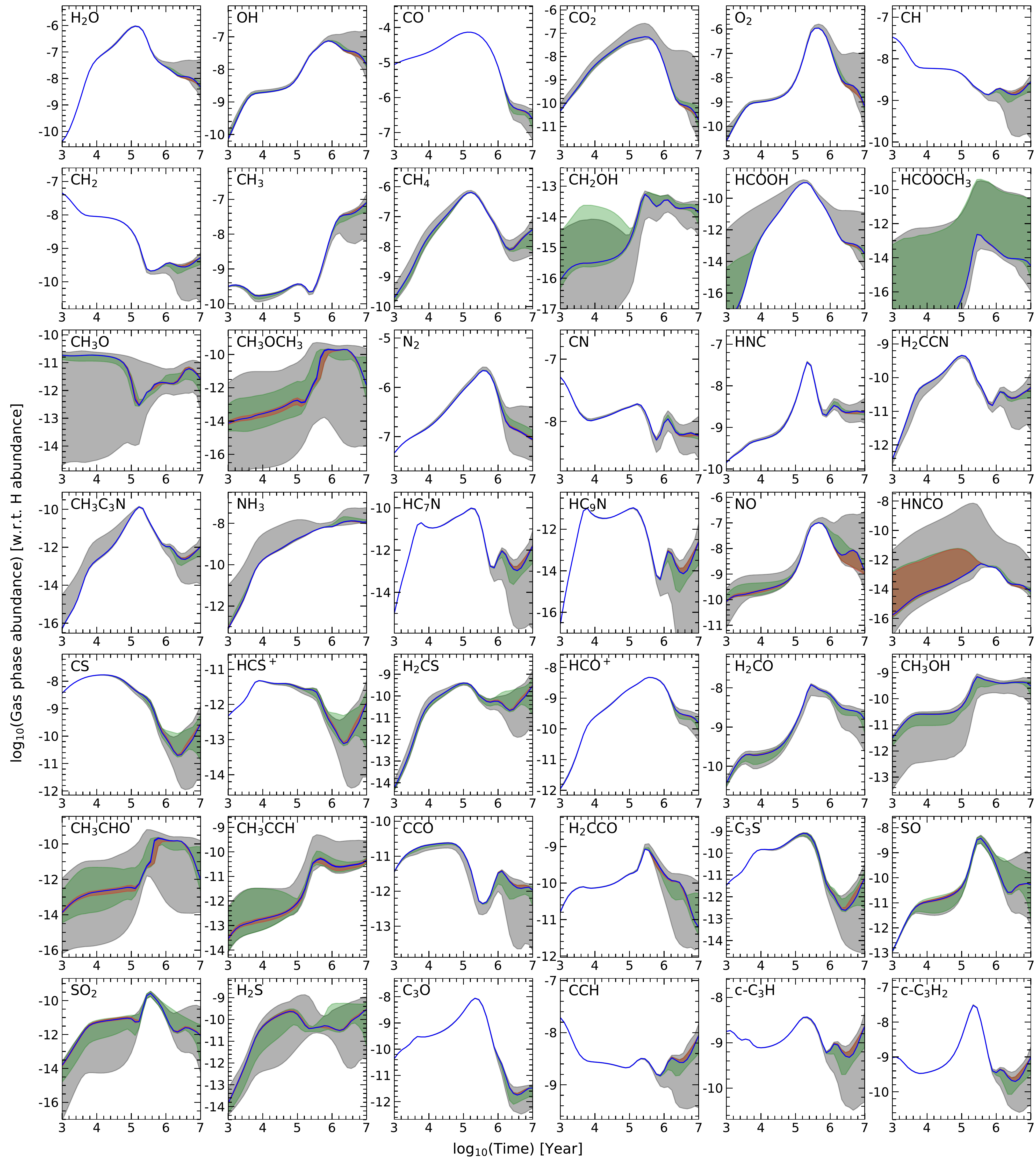}
   \caption{Gas-phase abundances within 2 $\sigma$  variation for different cases as a function of time. The area below the gray curve shows the variation for case A, the area below  the green curve is for case B, the area below the red curve is for case C, and the area below the blue curve is for case D; see Table \ref{tbl:case}.}
   \label{Fig:4}
\end{figure*}
\begin{figure}
    \centering
   \includegraphics[width=.5\textwidth,trim = 0cm 0cm 0cm 0cm, clip,angle=0]{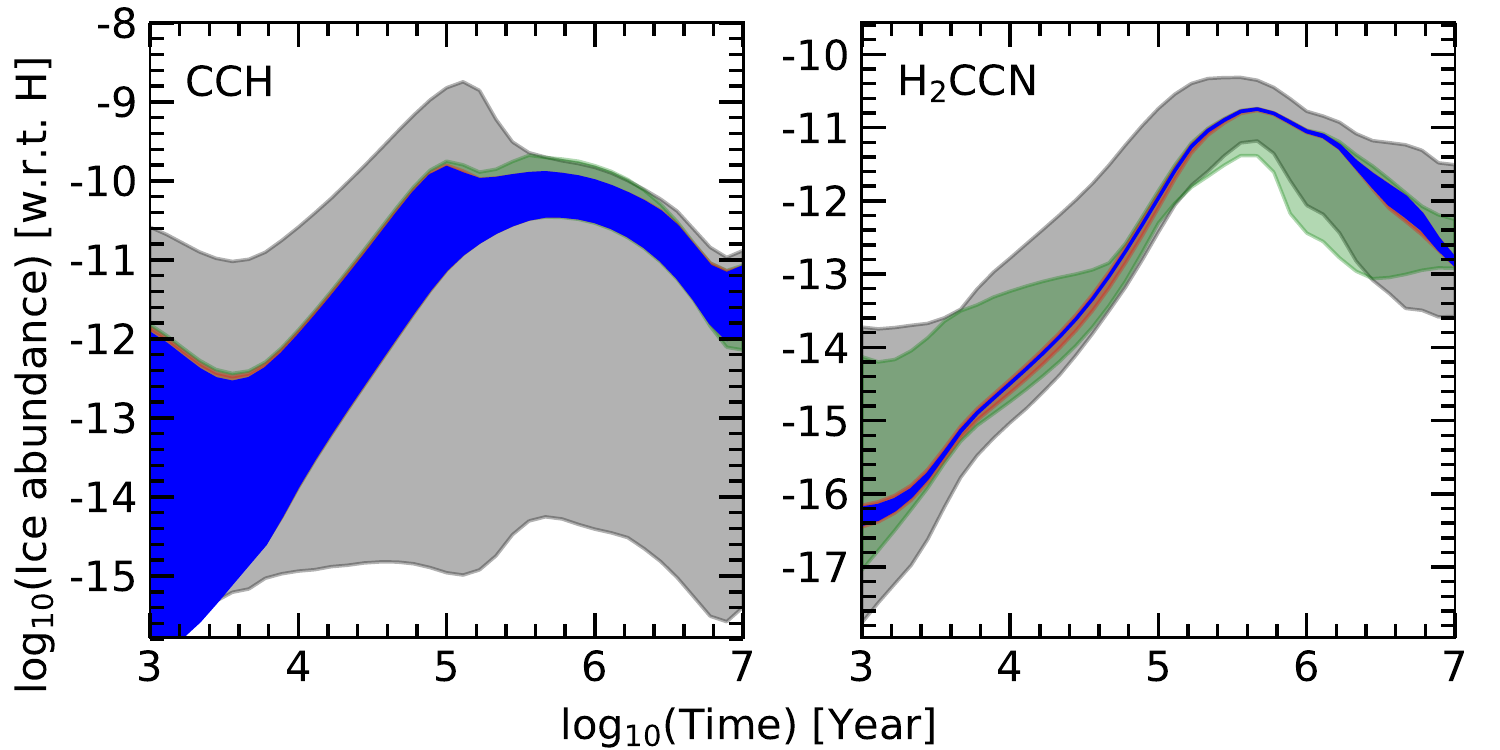}
   \caption{Same as Fig. \ref{Fig:4}, but for the ice abundances of CCH and H$_2$CCN alone.}
   \label{Fig:5}
\end{figure}
%
%
One of our objectives here is to identify the key species (if possible) for which a better estimate of $\mu^s$ would greatly reduce the uncertainties in the computed abundances. For this, we calculated Pearson correlation coefficients ($P(\mu^s,X(t))$) \citep[see][]{Penteado2017} between $\mu^s$  values of 86 key surface species (such as water, CO$_2$ and other  species observed in dark clouds) and their abundances in the gas phase and in the ice. 
Briefly, a Pearson correlation coefficient is a number between -1 and 1 that indicates the extent to which two variables are linearly related. A value of 1 or -1 means a strong correlation and anticorrelation, respectively, while a value of 0 means no correlation at all. Thus in our case, $P(\mu^s,X(t))$ gives a measure of the linear correlation or anticorrelation between $\mu^s$ values and the ice abundances of selected species.

On the basis of $P(\mu^s,X(t))$, we divided all species into four groups. In the first group we have species with absolute values of $P(\mu^s,X(t))$ equal to or greater than 0.3 for most of the species. H$_2$ ($E_b =$ 440K, \cite{Wakelam2017}), N ($E_b =$ 800K, \cite{Tielens1982}),O ($E_b =$ 800K, \cite{Tielens1987}), CO ($E_b =$ 1150K, OSU database), and CH$_3$ ($E_b =$ 1175K, OSU database) are the species in this group. In the second group we have (CH$_2$ ($E_b = $1050K, OSU database), HCO ($E_b =$ 1600K, OSU database), S ($E_b =$ 1100K, OSU database), and O$_2$ ($E_b =$ 1000K, OSU database) in decreasing order of the number of species that they are correlated or anticorrelated with) species with the absolute value of $P(\mu^s,X(t))$ equal to or greater than 0.3 for a considerable number of the species, but a weak correlation. In the third group we have NO ($E_b =$ 1600K, OSU database), HS ($E_b =$ 1450K, OSU database), and CH ($E_b =$ 925K, OSU database). These species are correlated or anticorrelated with only a few species.  In the last group we have all other species. The  $\mu^s$ values of these species show no correlation or anticorrelation with the abundances of other species.
Although carbon is the fourth most abundant species (see Table \ref{tbl:initial-abundances}), it is to be noted that the diffusion of C atoms does not introduce any noticeable uncertainties in the ice abundances of any species. This is because in our model the $E_b$ (= 4000K) value of the C atom is taken from \cite{Ruaud2015},  and \cite{Wakelam2017} also found similar values for the carbon atom. Such a high value of $E_b$ makes diffusion of carbon atoms inefficient.

$P(\mu^s,X(t))$ value gives a good idea about which species to look for giving maximum variation in the abundances of any particular species, but most often, this information alone is not enough to find the reactions that cause the variation in the species abundance. For example, in Fig. \ref{Fig:1} we see the large variation in the abundance of HNCO in both the gas phase and on ice, but we find no strong correlation with any surface species (see Fig. \ref{Fig:3}). The HNCO molecule is predominantly formed on the grain surface through the reaction
\begin{equation*}
\rm JH + JOCN \rightarrow JHNCO.
\end{equation*}
The chemical desorption during this reaction is responsible for the HNCO in the gas phase. The OCN on the grains mostly comes from the adsorption of gas-phase OCN, which is correlated to CO, N, O, and CH$_3$ (see Fig. \ref{Fig:3}). Gas-phase OCN is formed by
\begin{equation*}
\rm N + HCO \rightarrow H + OCN,
\end{equation*}
while HCO in the gas comes from the hydrogenation of CO on the grains. Thus, based on all these reactions, we can determine how the variation in $\mu^s$ of CO propagates to HNCO.
 
Now, we show three cases (cases B, C, and D, see Table \ref{tbl:case}) with 2000 new simulations in each case and compare it with case A, which we have discussed in the previous section. To recall, in case A we varied $\mu^s$ for all species except for H. In comparison, in case B we kept $\mu^s$ constant at 0.4 for all the species in group 1 (H$_2$, N, O, CO, and CH$_3$) in all 2000 simulations and varied $\mu^s$ for all other species except for H. In case C we kept $\mu^s$ constant at 0.4 for all the species in the second group (H$_2$, N, O, CO, CH$_3$, CH$_2$, HCO, S, and O$_2$) in all 2000 simulations and varied $\mu^s$ for all other species except for H. In case D we kept $\mu^s$ constant at 0.4 for H$_2$, N, O, CO, CH$_3$, CH$_2$, HCO, S, O$_2$, NO, HS, and CH in all 2000 simulations and varied $\mu^s$ for all other species except for H. 

In Fig. \ref{Fig:4} we plot the gas-phase abundances within 2 $\sigma$ variation for 42 key species. The figure clearly shows the effect of restricting the $\mu^s$ values of species in the sorted groups and how they affect the gas-phase abundances of other species. By comparing the areas below the gray and green curves, we see that uncertainties in the gas-phase abundances of most of the simple species, OH, CO, CN, N$_2$, NO, NH$_3$, H$_2$O, CO$_2$ etc., and some complex species, CH$_3$O, CH$_3$OH etc., can be reduced to just a very small fraction of the actual uncertainties by just restricting the uncertainties in the mobilities of five key species (H$_2$, N, O, CO, and CH$_3$). Furthermore, the area below the red curve shows that if we can do the same for other four species (CH$_2$, HCO, S, and O$_2$), then the uncertainties in the gas-phase abundances are reduced significantly for almost all the species except for a few, which depend strongly on the mobility of mostly NO in the ice. We therefore further limited the $\mu^s$ values of NO, HS, and CH and found that there are almost no uncertainties in the gas-phase abundances of any species because the blue curve appears just like a thin line. We did not find a noticeable uncertainty in the ice abundances either, with only a few species as exceptions. We found a few species in the network that strongly depend on the mobility of other species such as CN or C$_2$ . To show this, we plot the ice abundances of CCH and H$_2$CCN in Fig. \ref{Fig:5}. Even after restricting the values of $\mu^s$ for all identified species, we have a huge uncertainty in the ice abundance of CCH and a small but visible uncertainty in the ice abundance of H$_2$CCN. Basically, the huge uncertainty in the ice abundance of CCH comes from the uncertainty in $\mu^s$ of C$_2$ as CCH is formed in the ice through the reaction
\begin{equation*}
\rm JH + JC_2 \rightarrow JCCH,
\end{equation*}
and the uncertainty in the ice abundance of H$_2$CCN comes from the uncertainty in $\mu^s$ of CN through the reaction
\begin{equation*}
\rm JCH_2 + JCN \rightarrow H_2CNN.
\end{equation*}

The results shown above clearly suggest that the uncertainties (as far as the uncertainty in the chemical network that is due to the mobility of adsorbed species is concerned) in the abundances (both the gas phase and the ice) of most of the species can be very much removed by fixing the value of $\mu^s$ for 13 species, including H. Some species still are an exception and can be treated on a case basis if needed in the future.
\subsection{Effect of the uncertainties in $\mu^s$ of individual species}
\label{R_sec:3}
\begin{figure}
    \centering
   \includegraphics[width=0.5\textwidth,trim = 0cm 0cm 0cm 0cm, clip,angle=0]{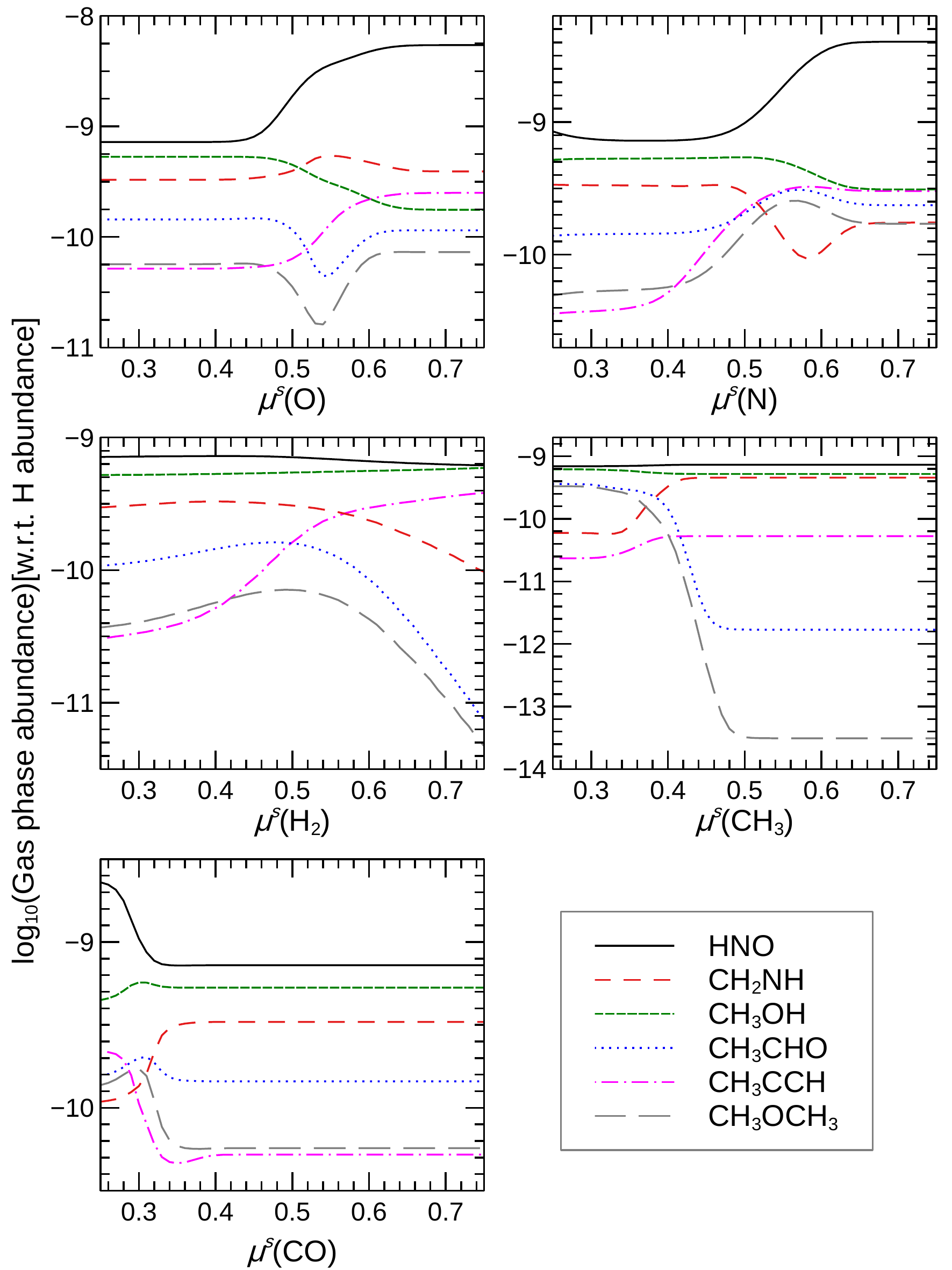}
   \caption{Calculated gas phase abundances of selected species at $4.6\times 10^5$ years as a function of $\mu^s({\rm X})$, where X = CO, N, O, H$_2$ or CH$_3$. $\mu^s$ of all other species were constant at 0.4. Legends apply to all plots.}
   \label{Fig:7}
\end{figure}
\begin{figure}
     \centering
    \includegraphics[width=0.5\textwidth,trim = 0cm 0cm 0cm 0cm, clip,angle=0]{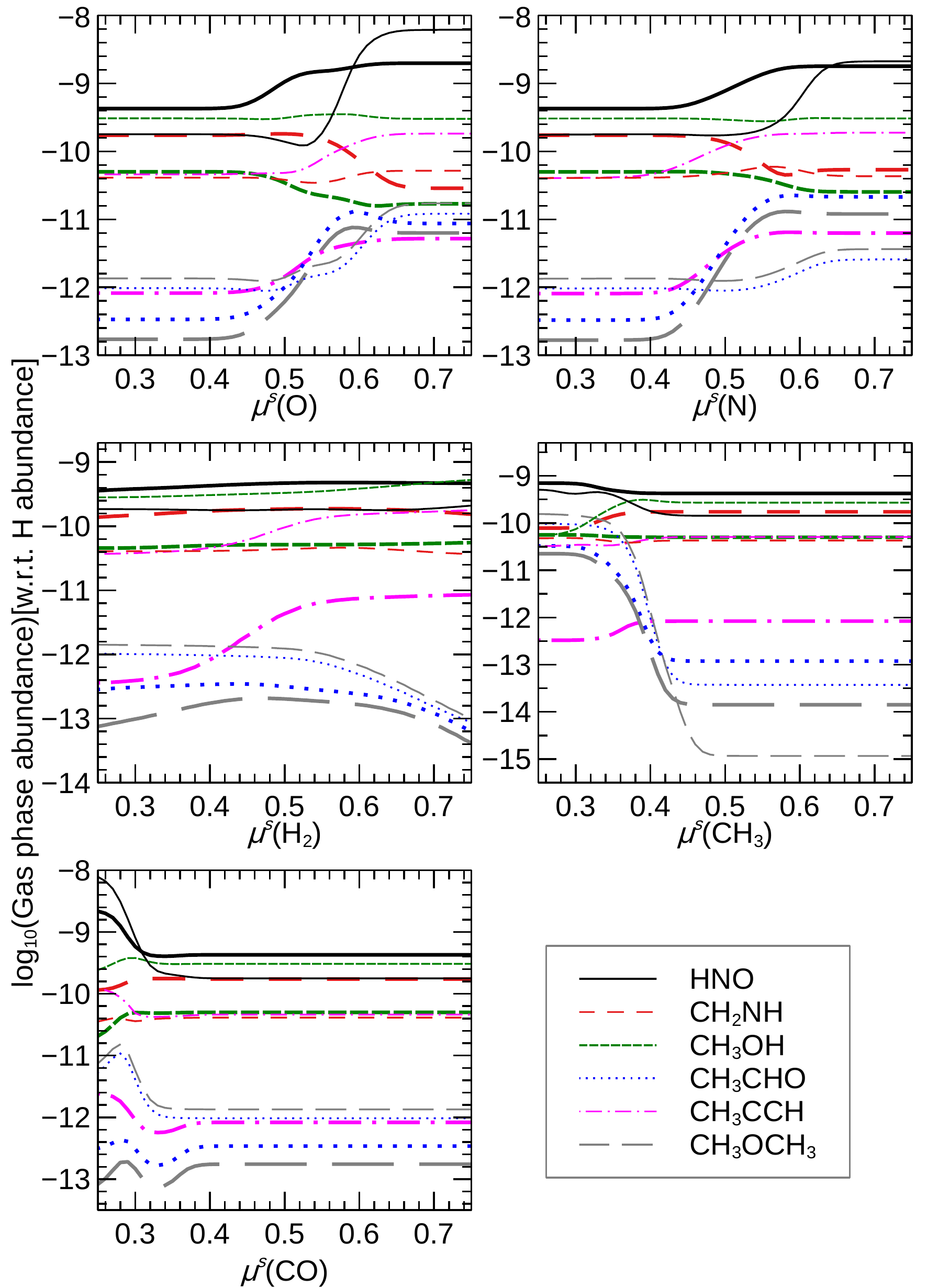}
    \caption{Same as Fig. \ref{Fig:7}, but for $10^4$ (thick lines) and $10^7$ (thin lines) years.}
    \label{Fig:8}
 \end{figure}

In this section we evaluate how the uncertainties in the values of $\mu^s$ of each of the species H$_2$, N, O, CO, and CH$_3$ individually affect the evolution of the chemical network. For this we ran 51 simulations in which we varied the values of $\mu^s$ for one of the species H$_2$, N, O, CO, and CH$_3$. Thus we ran five sets of simulations with 51 simulations in each set (one for each species). In the first set of simulations, we varied the value of $\mu^s$ of CH$_3$ between 0.25 to 0.75 with an increase of 0.1 in each subsequent simulation and kept it constant at 0.4 for all other species. In the following sets of simulations, we repeated this for the other species listed above. In this way, we can determine the effect of the mobility of one single species. 

In Fig. \ref{Fig:7} we plot the gas-phase abundances of six species (these species show high sensitivity to $\mu^s$ values of the selected species in case B) at $4.6 \times 10^5$ years as a function of $\mu^s$ of the selected species. In all plots, we see a common trend that there is a certain range of $\mu^s$ within which any change in the value of $\mu^s$ results in noticeable changes in the species abundances. In the case of CO, we see that when $\mu^s$ varies between 0.25 and 0.35, there are rapid changes in the abundances of the plotted species, but for any value of $\mu^s$ above 0.35, we obtain flat lines. For O we see the same range, but when $\mu^s$ is between 0.45 and 0.6. For N, this range is between 0.4 to 0.65, and for CH$_3$ this range is between 0.3 and 0.5. H$_2$ seems to be an exception as we see that any change in its $\mu^s$ value changes the abundance of plotted species. 
The very strong sensitivity to the H$_2$ diffusion is due to the reaction-diffusion competition included in the model, which strongly increases the surface reactions with H$_2$.

Furthermore, to verify how this range changes at other times in the simulations, we replotted the same plots, but at $10^4$ and $10^7$ in Fig. \ref{Fig:8}. We find that the sensitivity range of $\mu^s$ remains very much the same at different times in the simulation, but as expected, the abundances within the range change quite strongly with time. Thus we can conclude that the abundance of any species is sensitive only within a certain range of $\mu^s$ values of any surface species, and this range remains very much the same throughout the simulation. 

Here we also note if the diffusion timescale of a species becomes shorter than the timescale of the monolayer ice formation, then the diffusion of this species would be unimportant. We can define a critical diffusion energy where the two timescales become comparable. In our simulations, at $10^5$ years, ice is formed at a rate of about $4.9\times10^{-4}$ monolayers per year, which gives a critical diffusion energy of 532K. This means that the chemical composition should not be much influenced by species with diffusion energies higher than 532K. This critical energy becomes slightly higher with time (as the ice formation slows down): 558K and 589 at $10^6$ and $10^7$ years, respectively. 


In Sec. \ref{R_sec:1}, when we randomly changed $\mu^s$ of all species, we obtained large uncertainties in the species abundances, and then in Sec. \ref{R_sec:2}, we saw that the greater portion of the uncertainties in the species abundances comes from the uncertainties in the $\mu^s$ values of only a few surface species. To quantify the contribution of H$_2$, N, O, CO, and CH$_3$  to the total uncertainty in the gas-phase abundances of 80 key species, of which about 60 are observed in TMC-1 (CP), we calculated the total variation in the abundances of these species at $4.6 \times 10^5$ and $4.6 \times 10^6$ years in each of the five sets of simulations discussed above. This variation gives us the uncertainty that is due to the values of $\mu^s$ for H$_2$, N, O, CO, and CH$_3$ individually. We also calculated the total uncertainties in the gas-phase abundances of these 80 species in all 10000 simulation discussed in Sec. \ref{R_sec:1}. We list all these values in Table \ref{tbl:4}. In the first column of the table we list all species. In the next subsequent columns we list the uncertainties in the species abundances. Thus the column with the header $\mu^s{\rm (X)}$ means that the column contains uncertainties that are due to the value of $\mu^s$ of surface species X alone, while $\mu^s$ of other species was constant at 0.4. These values are in log of base 10, thus an uncertainty of 1 means an uncertainty of order one in the calculated abundance. 
From this table, we can find the source of the uncertainty in the abundance of any species in greater detail. Here it should be noted that the uncertainties from different sources cannot be summed directly to derive the total uncertainty because in the simulations these species are connected through various chemical networks. When we introduce uncertainties in all of them, then some add and some cancel, and some show other possible outcomes. We still found that a simple addition gives a rough but good estimate of the total uncertainty introduced by selected species, however.

\subsection{Sensitivity to the surface temperature}
\label{R_sec:4}
To determine how the uncertainty in the abundance of any species varies with the change in surface temperature, we ran case A again, but this time, we ran 2000 simulations with a surface temperature at 8K and another 2000 simulations with a surface temperature at 12K. The surface temperature was kept at 10K in the first case A presented in Sec. \ref{R_sec:1}. We found that the species that are predominantly formed in the gas phase do not show any significant change in the total uncertainties with a change in temperature. Only the mean value of the abundance changes. The species predominantly formed on the grain surface, in contrast, showed noticeable changes in the total uncertainty, but these changes are not unidirectional. For example, species such as CH$_2$OH, HCOOH, HNCO, CS, H$_2$S, and CH$_3$CCH show a jump in the total uncertainty at 12K and a reduction in the total uncertainty at 8K. Other species such as CH$_3$O, N$_2$, HNC, NH$_3$, and CH$_3$OH show a completely opposite trend.

\section{Best model and comparison with observations}
\label{R_sec:5}
\begin{figure}
     \centering
    \includegraphics[width=0.5\textwidth,trim = 0cm 0cm 0cm 0cm, clip,angle=0]{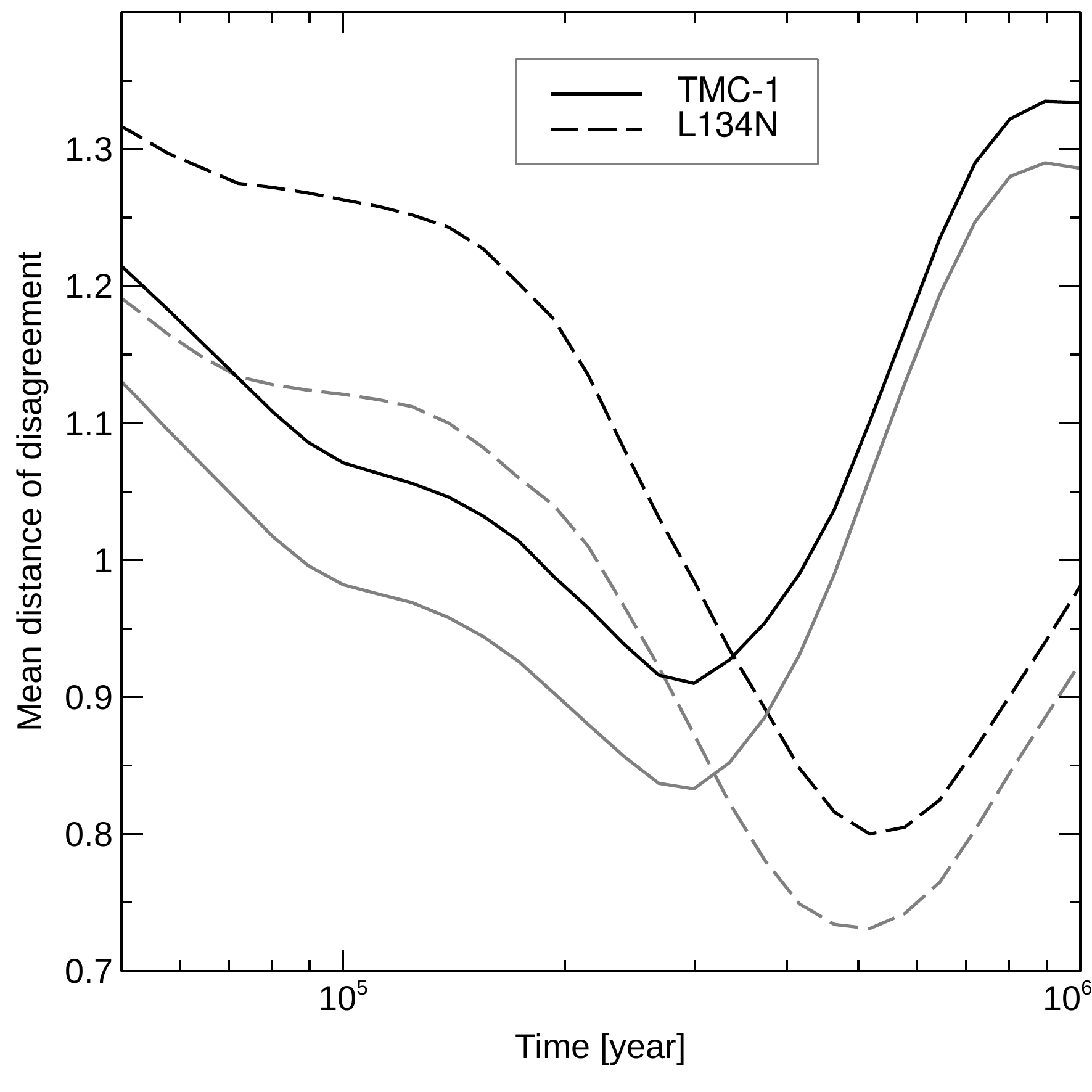}
    \caption{Mean distance of disagreement as a function of time for different models. The black lines show the model in which $\mu^s = 0.4$ for all species, and the gray lines show the model with best-fit values of $\mu^s$ for H$_2$, CH$_3$ , and CO.}
    \label{Fig:9}
 \end{figure}
\begin{table*}
   \caption{Best-fit (at best-fit time of 2.99$\times 10^5$ years) ice composition against observations in MYSOs, LYSOs, and toward BG stars. Observed values (in \% of H$_2$O ice) are taken from \citet{Boogert2015}.}
   \label{tbl:ice_obs}
   \centering
   \begin{tabular}{llllllllll}
   \hline
   \hline
   Model/Observations                      & H$_2$O  & CO     & CO$_2 $  & CH$_3$OH  & NH$_3$  & CH$_4$  & H$_2$CO       & OCS    \\
   \hline                                                                                                                                                
   Best-fit model                          & 100     & 10.74  & 44.04    & 10.02     & 1.52    & 11.70   & 6.41          & 0.04   \\
   Nautilus (with $\mu^s=0.4$)             & 100     & 13.76  & 54.77    & 12.56     & 1.10    & 10.10   & 7.63          & 0.02   \\
   \hline                                                                                                
   BG stars                                &100      & $20-43$  & $18-39$    & $6-10$      & $<7$      & $ <3$     & nd           & $<0.22$     \\
   MYSOs                                   &100      & $4-15$   & $12-25$    & $5-23$      & $\sim7$   & $1-3  $   & $\sim2-7$    & $0.03-0.16$ \\
   LYSOs                                   &100      & $12-35$  & $23-37$    & $5-12$      & $4-8$     & $3-6$     & $\sim 6$     & $\leq1.6 $  \\
   \hline
   \end{tabular}
\end{table*}

We used the simple method of parameter fitting to best fit the observed abundances of 58 species, observed in the dark cloud TMC-1 (CP). We used the observed abundances gathered in \citet{Agundez2013} for cold cores. In our calculations, we did not use those species for which only the upper or the lower limits of the abundance is given. Knowing that most of the uncertainty in the simulated abundance comes from only a few species, we determined the best value of $\mu^s$ for H$_2$, CH$_3$, and CO to obtain a better agreement with the observed abundances. Of course, this method assumes that most of the error in the model is mainly due to the uncertainty in $\mu^s$. 

For comparison with observed abundances, we used the method described in \cite{Loison2014}. We computed the mean distance of disagreement $D(t)$ for each output of the simulation using the formula
\begin{equation}
 D(t) = \frac{1}{N_{obs}} \sum_{i}|{\rm log_{10}}[n(X_i,t)]-{\rm log_{10}}[n(X_i^{obs})]|
,\end{equation}
where $n(X_i,t)$ is the calculated abundance of species $X_i$  at time $t$ , $n(X_i^{obs})$  is the observed abundance for the same species, and $N_{obs}$ is the total number of observed species used in the calculation of $D(t)$. In this method, the model with the lower value of $D(t)$ is the best model because a lower value of $D(t)$  means a better agreement between the observed abundances and the simulated results. 

To calculate the best-fit values of $\mu^s$, we used an iterative method. In this method we fixed $\mu^s$ at 0.4 for all species except for CO. Then we ran 51 simulations in which we varied $\mu^s({\rm CO})$ between 0.25 to 0.75. Then we compared abundances in all 51 simulations with the observed abundances in TMC-1 (CP) and determined which value of $\mu^s({\rm CO})$ gives the best agreement. Next we fixed $\mu^s({\rm CO})$ at this value and repeated the same process for CH$_3$.
We then fixed $\mu^s({\rm CH_3})$ to its best value and repeated this again for CO. 
We repeated the same process many times until we converged to the best values of $\mu^s({\rm CO})$ and $\mu^s({\rm CH_3})$, which did not change any more. Then we introduced the third species H$_2$ and used the same iterative process to determine the best values of $\mu^s({\rm CO})$, $\mu^s({\rm CH_3})$, and $\mu^s({\rm H_2})$. The best-fit values are 0.33, 0.34, and 0.74, corresponding to diffusion energies of 380K, 399K, and 326K for CO, CH$_3$ , and H$_2$ , respectively. Next, we ran a simulation in which we used the bes- fit values for $\mu^s({\rm CO})$, $\mu^s({\rm CH_3})$, and $\mu^s({\rm H_2})$ and kept $\mu^s$ = 0.4 for all other species. For comparison, we also ran one simulation in which we kept $\mu^s$ = 0.4 for all species. 

In our iterative method for calculating the best-fit values, we only used the observed abundances in TMC-1 (CP). This method does not guarantee that if we use observational data of another cloud, we would also obtain the same best values of $\mu^s$. Nevertheless, we also calculated $D(T)$ for 34 observed molecules for an other cold core, L134N (N). Again, for observational data for L134N, we used the tabulated values in \citet{Agundez2013}.

In Fig. \ref{Fig:9} we show the calculated $D(t)$ values for both the best-fit and the standard models and for both clouds. The best-fit model improves the agreement with the observed abundances for both clouds. This improvement is of about 10\% in log$_{10}$ scale (0.1dex); this is closer to a 25\% improvement in the linear scale. Furthermore, this is an average or overall improvement in agreement with the observational values of 58 and 34 observed species in TMC-1 (CP) and L134N (N), respectively. When we consider only certain species, then the improvement is quite significant. The best-fit model gives a better agreement for 32 out of 58 species in TMC-1 (CP). For these 32 species, the lowest value of $D(t)$ is 0.67 for the best-fit model, while it is 0.86 for the standard model, which is an overall improvement of 22\% (in log$_{10}$ scale). For the other 26 species, the lowest value of $D(t)$ is 1.04 for the best-fit model, while it is 0.98 for the standard model. This is a negative change of only 6\% (in log$_{10}$ scale) for the best-fit model.

In Table \ref{tbl:ice_obs} we show the comparison between the best-fit model, the standard Nautilus model (with $\mu^s = .4$), and the ice abundances observed in the envelopes around young stellar objects (LYSOs), massive young stellar objects (MYSOs), and toward the background stars (BG stars) as given in \cite{Boogert2015}. For comparison with simulated results, we used the values at best-fit times (2.99 $\times 10^5$ years). It is to be noted that our simulated results are for dark cloud conditions and do not properly represent the conditions in LYSOs, MYSOs, or the regions toward the BG stars. Nevertheless, for a qualitative study it is good to compare different models with observations. Table \ref{tbl:ice_obs} shows that the two models are not much different in  their results: we observe only a slightly better agreement for the new model. However, both CO$_2$ and CH$_4$ are overproduced by the models.

%
%
%
%

%
%
%
%
%
\section{Conclusions}
We calculated the uncertainties induced in the gas phase and the ice abundances of the species  in simulated results due to the uncertainties in the diffusion energy of adsorbed species. We showed that these uncertainties are very large for species that are predominantly formed on the grain surface, such as CH$_3$OH. Species such as CO, OH, or HNC, which are mainly formed in the gas phase, only show an uncertainty after $5 \times 10^5$ years in simulations. By this time, we have more than 100 monolayers of ice on the grain surface. This large reservoir of species on the grain surface causes the propagation of uncertainties in the gas-phase abundances. 

We identified the surface species, the uncertainties in diffusion rates of which result in the larger variations in abundances of most of the species.  CO, H$_2$, O, N, and CH$_3$ are the key species for the uncertainties in the abundances, while CH$_2$, HCO, S, and O$_2$ come next, followed by NO, HS, and CH. We also showed that by limiting the uncertainties in the  ratio of  diffusion to binding energy of these species, we were able to eliminate the uncertainties in the gas-phase abundances of almost all the species.

We calculated the contribution to the uncertainties in the species abundances  coming from CO, H$_2$, O, N, and CH$_3$ individually. We found that there is a small range of $\mu^s$ (the ratio of diffusion energy to binding energy) for all these species within which we obtain variation in the abundances. The size and position of this range varies with species. For O, it is between 0.45 to 0.6. For N, it is 0.4 to 0.65. For CO, it is 0.25 to 0.36. For CH$_3$, it is 0.36 to 0.5. H$_2$ is an exception and affects the surface chemistry within the entire range of $\mu^s$ we tested in our simulations.  We found that this range remains constant throughout the simulation and thus does not depend on time. 

We also tested for different grain surface temperatures and found that although a different surface temperature changes the uncertainty pattern for most of the species, it does not follow a unidirectional increase or decrease in the variation in the species abundances. Thus it is not possible to predict if the uncertainty in the abundance of a certain species will increase or decrease with an increase or decrease of the surface temperature. 

We calculated the best values of $\mu^s$ for CO, CH$_3$ , and H$_2$, using an iterative approach, to better explain the observed abundances in dark dense cloud such as TMC-1(CP). We showed that we can improve the overall agreement by 8\% (in log$_{10}$ scale) for 58 species and 22\% (in log$_{10}$ scale) for 32 species observed in TMC-1 (CP) by just fixing the $\mu^s$ values of CO, CH$_3$ , and H$_2$ at 0.33, 0.34, and 0.74, respectively.  

We provided uncertainties in abundances of a number of species due to individual species (H$_2$, N, O, CO, and CH$_3$) and also cumulative uncertainties due to all surface species in a tabulated form. This table provides an idea of how much error one can expect in their modeling of observed abundances in similar cold cores. 

It should be noted that these values and other results presented in this work may depend on the physical conditions of the clouds and also on the model parameters. The surface temperature alone can significantly increase or decrease the uncertainties.   

In our simulations, we kept the binding energies of all species constant and only varied the diffusion energy depending on values of $\mu^s$. The key species  we found are not expected to change in a different set of binding energies unless the difference is very large (hundreds of K). The range of $\mu^s$ calculated for CO, O, N, and CH$_3$ is expected to shift by a few points left or right until the diffusion energy matches the energy we used in our simulation.
\begin{acknowledgements}
This study has received financial support from the French State  in  the  frame  of  the  "Investments  for  the  future"  Programme  IdEx  Bordeaux, reference ANR-10-IDEX-03-02. The research of VW is funded by an ERC Starting Grant (3DICE, grant agreement 336474) and the CNRS program Physique et Chimie du Milieu Interstellaire (PCMI), co-funded by the Centre National d'Etudes Spatiales (CNES).
\end{acknowledgements}
%
%
%
%
%
%
\bibliographystyle{aa} 
\bibliography{Pv7}    

\longtab{
\begin{longtable}{lllllll|llllll}
   \caption{\label{tbl:4} Summary of different cases.}\\
   
   \hline
   \hline
   Species     & $\mu^s({\rm O})$    &$\mu^s({\rm N})$ &  $\mu^s({\rm H_2})$ &  $\mu^s({\rm CO})$   & $\mu^s({\rm CH_3})$ & $\mu^s({\rm All})$ &               $\mu^s({\rm O})$ &  $\mu^s({\rm N})$ &  $\mu^s({\rm H_2})$ & $\mu^s({\rm CO})$ & $\mu^s({\rm CH_3})$ & $\mu^s({\rm All})$ \\    
 \hline   
              &            \multicolumn{6}{c|}{At $4.6 \times 10^5$ years}                                     &                       \multicolumn{6}{c}{At $4.6 \times 10^6$ years}                                    \\
 \hline          
 \endfirsthead
\caption{continued.}\\
\hline\hline
   Species     & $\mu^s({\rm O})$    &$\mu^s({\rm N})$ &  $\mu^s({\rm H_2})$ &  $\mu^s({\rm CO})  $ & $\mu^s({\rm CH_3})$ & $\mu^s({\rm All})$ &               $\mu^s({\rm O})$ &  $\mu^s({\rm N})$ &  $\mu^s({\rm H_2})$ & $\mu^s({\rm CO})$ & $\mu^s({\rm CH_3})$ & $\mu^s({\rm All})$ \\    
  \hline
               &            \multicolumn{6}{c|}{At $5\times10^5$ years}                                     &                       \multicolumn{6}{c}{At $5\times10^6$ years}                                    \\
 \hline          
\endhead
\hline
\endfoot
 O$_2$               &      0.11     &     0.05  &     0.01      &      0.05    &    0.00         &       0.17   &                    0.84  &       0.71  &      0.16     &     1.07    &     0.47       &       1.68  \\
 H$_2$O              &      0.04     &     0.02  &     0.00      &      0.02    &    0.00         &       0.07   &                    0.36  &       0.33  &      0.06     &     0.51    &     0.23       &       0.78  \\
 OH                  &      0.05     &     0.02  &     0.00      &      0.02    &    0.00         &       0.06   &                    0.37  &       0.33  &      0.07     &     0.51    &     0.23       &       0.78  \\
 CO                  &      0.00     &     0.00  &     0.00      &      0.00    &    0.00         &       0.00   &                    0.08  &       0.02  &      0.08     &     0.27    &     0.40       &       0.53  \\
 CO$_2$              &      0.36     &     0.03  &     0.01      &      0.05    &    0.00         &       0.55   &                    1.35  &       0.39  &      0.13     &     0.37    &     0.37       &       2.62  \\
 CH                  &      0.07     &     0.05  &     0.00      &      0.03    &    0.00         &       0.09   &                    0.43  &       0.38  &      0.08     &     0.82    &     0.26       &       1.17  \\
 CH$_2$              &      0.09     &     0.05  &     0.00      &      0.04    &    0.00         &       0.11   &                    0.43  &       0.38  &      0.08     &     0.81    &     0.26       &       1.17  \\
 CH$_3$              &      0.08     &     0.04  &     0.01      &      0.03    &    0.02         &       0.14   &                    0.38  &       0.31  &      0.13     &     0.64    &     0.20       &       1.01  \\
 CH$_4$              &      0.02     &     0.01  &     0.01      &      0.01    &    0.00         &       0.09   &                    0.45  &       0.29  &      0.42     &     0.30    &     0.45       &       0.85  \\
 CCO                 &      0.28     &     0.15  &     0.01      &      0.15    &    0.02         &       0.33   &                    0.50  &       0.45  &      0.07     &     1.27    &     0.97       &       1.95  \\
 C$_3$O              &      0.00     &     0.02  &     0.00      &      0.00    &    0.00         &       0.03   &                    0.45  &       0.31  &      0.08     &     0.83    &     0.47       &       1.05  \\
 CCH                 &      0.03     &     0.04  &     0.00      &      0.01    &    0.00         &       0.06   &                    0.49  &       0.40  &      0.14     &     0.83    &     0.48       &       1.22  \\
 C$_2$H$_2$          &      0.05     &     0.03  &     0.01      &      0.02    &    0.01         &       0.07   &                    0.49  &       0.41  &      0.13     &     0.85    &     0.46       &       1.19  \\
 C$_4$H              &      0.05     &     0.05  &     0.00      &      0.02    &    0.00         &       0.10   &                    0.96  &       0.78  &      0.17     &     1.70    &     0.75       &       2.20  \\
 C$_5$H              &      0.09     &     0.07  &     0.00      &      0.04    &    0.00         &       0.15   &                    1.20  &       1.00  &      0.22     &     2.21    &     0.92       &       2.88  \\
 C$_4$H$_2$          &      0.00     &     0.03  &     0.01      &      0.01    &    0.00         &       0.04   &                    0.80  &       0.63  &      0.14     &     1.41    &     0.74       &       1.90  \\
 C$_6$H              &      0.12     &     0.08  &     0.01      &      0.05    &    0.01         &       0.20   &                    1.46  &       1.22  &      0.23     &     2.74    &     1.05       &       3.51  \\
 C$_6$H$^-$          &      0.13     &     0.09  &     0.01      &      0.05    &    0.01         &       0.22   &                    1.49  &       1.25  &      0.23     &     2.79    &     1.07       &       3.56  \\
 C$_6$H$_2$          &      0.09     &     0.07  &     0.01      &      0.04    &    0.01         &       0.18   &                    1.38  &       1.14  &      0.23     &     2.55    &     1.04       &       3.28  \\
 C$_8$H              &      0.09     &     0.07  &     0.01      &      0.04    &    0.01         &       0.15   &                    2.04  &       1.73  &      0.30     &     3.85    &     1.43       &       4.88  \\
 C$_8$H$^-$          &      0.10     &     0.08  &     0.01      &      0.04    &    0.01         &       0.17   &                    2.08  &       1.76  &      0.30     &     3.91    &     1.45       &       4.96  \\
 c-C$_3$H            &      0.03     &     0.04  &     0.00      &      0.01    &    0.00         &       0.06   &                    0.75  &       0.61  &      0.16     &     1.23    &     0.57       &       1.56  \\
 l-C$_3$H            &      0.02     &     0.03  &     0.00      &      0.01    &    0.00         &       0.04   &                    0.71  &       0.56  &      0.14     &     1.14    &     0.56       &       1.46  \\
 c-C$_3$H$_2$        &      0.00     &     0.01  &     0.00      &      0.00    &    0.00         &       0.02   &                    0.64  &       0.46  &      0.13     &     0.96    &     0.57       &       1.30  \\
 l-C$_3$H$_2$        &      0.00     &     0.01  &     0.00      &      0.00    &    0.00         &       0.02   &                    0.64  &       0.46  &      0.13     &     0.95    &     0.57       &       1.29  \\
 N                   &      0.02     &     0.04  &     0.00      &      0.01    &    0.00         &       0.07   &                    0.21  &       0.24  &      0.05     &     0.27    &     0.29       &       0.64  \\
 N$_2$               &      0.05     &     0.06  &     0.01      &      0.04    &    0.02         &       0.08   &                    0.14  &       0.33  &      0.07     &     0.42    &     0.30       &       1.03  \\
 NO                  &      0.16     &     0.11  &     0.00      &      0.06    &    0.00         &       0.18   &                    0.88  &       0.85  &      0.13     &     1.22    &     0.67       &       2.10  \\
 N$_2$O              &      0.20     &     0.54  &     0.00      &      0.13    &    0.01         &       0.58   &                    1.06  &       1.04  &      0.15     &     1.73    &     0.96       &       2.89  \\
 CN                  &      0.08     &     0.02  &     0.01      &      0.03    &    0.01         &       0.15   &                    0.04  &       0.13  &      0.08     &     0.22    &     0.15       &       0.49  \\
 CCN                 &      0.05     &     0.01  &     0.00      &      0.02    &    0.00         &       0.09   &                    0.30  &       0.30  &      0.17     &     0.63    &     0.46       &       1.06  \\
 C$_3$N              &      0.05     &     0.01  &     0.00      &      0.02    &    0.00         &       0.10   &                    0.51  &       0.45  &      0.19     &     0.95    &     0.57       &       1.29  \\
 C$_5$N              &      0.10     &     0.04  &     0.01      &      0.04    &    0.00         &       0.17   &                    0.87  &       0.72  &      0.23     &     1.68    &     0.88       &       2.32  \\
 HNO                 &      0.88     &     0.75  &     0.07      &      0.50    &    0.02         &       0.96   &                    0.73  &       0.72  &      0.22     &     0.96    &     0.60       &       2.23  \\
 HCO                 &      0.16     &     0.06  &     0.01      &      0.10    &    0.01         &       0.23   &                    0.08  &       0.07  &      0.10     &     0.39    &     0.89       &       0.89  \\
 HCN                 &      0.02     &     0.03  &     0.01      &      0.01    &    0.01         &       0.08   &                    0.03  &       0.16  &      0.09     &     0.15    &     0.21       &       0.44  \\
 HNC                 &      0.01     &     0.04  &     0.01      &      0.00    &    0.01         &       0.09   &                    0.03  &       0.16  &      0.08     &     0.15    &     0.20       &       0.44  \\
 HNCO                &      0.10     &     0.10  &     0.62      &      3.36    &    0.01         &       3.41   &                    0.68  &       0.62  &      0.11     &     2.98    &     0.59       &       2.78  \\
 HCCNC               &      0.01     &     0.02  &     0.00      &      0.01    &    0.01         &       0.07   &                    0.38  &       0.35  &      0.20     &     0.77    &     0.57       &       1.18  \\
 HNCCC               &      0.01     &     0.02  &     0.00      &      0.01    &    0.00         &       0.08   &                    0.41  &       0.37  &      0.20     &     0.82    &     0.57       &       1.23  \\
 NH$_3$              &      0.09     &     0.16  &     0.01      &      0.11    &    0.01         &       0.23   &                    0.23  &       0.29  &      0.07     &     0.56    &     0.36       &       0.77  \\
 NH$_2$CN            &      0.09     &     0.09  &     0.04      &      0.05    &    0.03         &       0.26   &                    0.15  &       0.09  &      0.05     &     0.32    &     0.09       &       0.40  \\
 HC$_7$N             &      0.07     &     0.02  &     0.00      &      0.03    &    0.00         &       0.15   &                    1.33  &       1.11  &      0.29     &     2.63    &     1.11       &       3.57  \\
 HC$_9$N             &      0.11     &     0.05  &     0.01      &      0.05    &    0.01         &       0.23   &                    1.90  &       1.60  &      0.36     &     3.72    &     1.42       &       4.95  \\
 SO                  &      0.16     &     0.30  &     0.03      &      0.15    &    0.02         &       0.38   &                    0.18  &       0.20  &      0.24     &     1.16    &     1.00       &       2.32  \\
 SO$_2$              &      0.11     &     0.28  &     0.04      &      0.14    &    0.01         &       0.37   &                    0.34  &       0.13  &      0.15     &     1.36    &     1.03       &       2.71  \\
 CS                  &      0.28     &     0.37  &     0.03      &      0.15    &    0.02         &       0.44   &                    0.77  &       0.63  &      0.13     &     0.40    &     0.67       &       2.32  \\
 CS$^+$              &      0.25     &     0.35  &     0.03      &      0.14    &    0.02         &       0.41   &                    0.97  &       0.81  &      0.14     &     0.70    &     0.71       &       2.49  \\
 NS                  &      0.40     &     0.42  &     0.06      &      0.20    &    0.11         &       1.36   &                    0.32  &       0.11  &      0.48     &     0.76    &     1.05       &       2.71  \\
 OCS                 &      0.20     &     0.42  &     0.05      &      0.16    &    0.06         &       0.55   &                    0.62  &       0.17  &      0.37     &     0.78    &     0.67       &       2.25  \\
 CCS                 &      0.30     &     0.37  &     0.03      &      0.15    &    0.02         &       0.49   &                    1.34  &       0.97  &      0.17     &     1.05    &     0.40       &       2.89  \\
 C$_3$S              &      0.27     &     0.36  &     0.03      &      0.15    &    0.02         &       0.44   &                    1.36  &       1.01  &      0.19     &     1.19    &     0.41       &       2.98  \\
 H$_2$S              &      0.27     &     0.18  &     0.02      &      0.37    &    0.09         &       0.85   &                    0.78  &       0.24  &      0.27     &     0.46    &     1.38       &       3.02  \\
 H$_2$CS             &      0.29     &     0.36  &     0.03      &      0.15    &    0.01         &       0.46   &                    0.87  &       0.57  &      0.13     &     0.32    &     0.75       &       2.36  \\
 CH$_2$OH            &      0.56     &     0.25  &     0.04      &      0.14    &    0.07         &       0.82   &                    0.14  &       0.18  &      0.09     &     0.54    &     0.50       &       0.89  \\
 CH$_3$O             &      0.55     &     0.29  &     0.11      &      0.23    &    0.33         &       1.32   &                    0.67  &       0.39  &      0.58     &     0.36    &     0.56       &       1.16  \\
 CH$_3$OH            &      0.48     &     0.24  &     0.05      &      0.11    &    0.07         &       0.66   &                    0.12  &       0.10  &      0.15     &     0.34    &     0.60       &       0.69  \\
 CH$_3$OCH$_3$       &      0.66     &     0.71  &     1.18      &      0.48    &    4.03         &       4.58   &                    0.52  &       0.47  &      1.87     &     0.58    &     5.40       &       6.10  \\
 HCOOH               &      0.06     &     0.03  &     0.01      &      0.27    &    0.00         &       0.28   &                    0.49  &       0.40  &      0.12     &     2.30    &     0.55       &       2.49  \\
 HCOOCH$_3$          &      0.60     &     0.31  &     0.08      &      0.15    &    0.10         &       3.99   &                    0.02  &       0.03  &      0.24     &     0.43    &     1.17       &       4.74  \\
 HCO$^+$             &      0.01     &     0.00  &     0.00      &      0.00    &    0.00         &       0.01   &                    0.10  &       0.04  &      0.08     &     0.25    &     0.38       &       0.52  \\
 H$_2$CO             &      0.18     &     0.10  &     0.02      &      0.07    &    0.01         &       0.23   &                    0.05  &       0.06  &      0.09     &     0.32    &     0.73       &       0.71  \\
 H$_2$CCO            &      0.06     &     0.10  &     0.10      &      0.04    &    0.20         &       0.29   &                    0.48  &       0.42  &      0.97     &     0.60    &     1.38       &       1.99  \\
 CH$_3$CHO           &      0.52     &     0.35  &     1.34      &      0.15    &    2.33         &       2.93   &                    0.54  &       0.47  &      1.87     &     0.60    &     3.41       &       4.29  \\
 CH$_3$CCH           &      0.69     &     0.96  &     1.10      &      0.68    &    0.35         &       1.25   &                    0.74  &       0.97  &      0.97     &     0.70    &     0.20       &       1.20  \\
 CH$_3$C$_6$H        &      0.05     &     0.04  &     0.02      &      0.03    &    0.00         &       0.08   &                    1.37  &       1.05  &      0.29     &     2.12    &     1.27       &       2.93  \\
 N$_2$H$^+$          &      0.06     &     0.07  &     0.01      &      0.05    &    0.02         &       0.08   &                    0.15  &       0.32  &      0.06     &     0.45    &     0.33       &       1.05  \\
 H$_2$CN             &      0.09     &     0.04  &     0.01      &      0.03    &    0.03         &       0.20   &                    0.18  &       0.21  &      0.16     &     0.37    &     0.19       &       0.75  \\
 HCNH$^+$            &      0.01     &     0.04  &     0.01      &      0.01    &    0.01         &       0.09   &                    0.02  &       0.15  &      0.08     &     0.17    &     0.20       &       0.45  \\
 H$_2$CCN            &      0.05     &     0.03  &     0.03      &      0.04    &    0.03         &       0.14   &                    0.30  &       0.30  &      0.19     &     0.69    &     0.20       &       1.22  \\
 CH$_3$CN            &      0.06     &     0.02  &     0.01      &      0.02    &    0.01         &       0.12   &                    0.29  &       0.31  &      0.13     &     0.78    &     0.18       &       1.28  \\
 CH$_2$CHCN          &      0.15     &     0.06  &     0.16      &      0.04    &    0.08         &       0.29   &                    0.21  &       0.21  &      0.87     &     0.16    &     1.18       &       2.01  \\
 HC$_3$N             &      0.01     &     0.02  &     0.00      &      0.01    &    0.01         &       0.08   &                    0.36  &       0.33  &      0.20     &     0.75    &     0.57       &       1.17  \\
 HC$_3$NH$^+$        &      0.01     &     0.02  &     0.00      &      0.01    &    0.00         &       0.08   &                    0.41  &       0.37  &      0.20     &     0.83    &     0.58       &       1.24  \\
 CH$_3$C$_3$N        &      0.05     &     0.02  &     0.00      &      0.02    &    0.01         &       0.11   &                    0.67  &       0.58  &      0.21     &     1.35    &     0.72       &       1.82  \\
 HC$_5$N             &      0.06     &     0.02  &     0.00      &      0.02    &    0.01         &       0.15   &                    0.77  &       0.62  &      0.21     &     1.50    &     0.83       &       2.16  \\
 CH$_3$C$_5$N        &      0.10     &     0.04  &     0.00      &      0.04    &    0.00         &       0.19   &                    1.05  &       0.88  &      0.26     &     2.16    &     0.90       &       3.04  \\
 CH$_3$COCH$_3$      &      0.90     &     1.00  &     1.41      &      0.40    &    3.02         &       3.75   &                    0.83  &       0.71  &      3.12     &     0.69    &     6.48       &       7.44  \\
 CH$_2$NH            &      0.22     &     0.55  &     0.53      &      0.48    &    0.90         &       1.24   &                    0.26  &       0.11  &      0.19     &     0.07    &     0.19       &       0.63  \\
 CH$_3$C$_4$H        &      0.00     &     0.02  &     0.01      &      0.01    &    0.00         &       0.10   &                    0.91  &       0.72  &      0.37     &     1.27    &     0.91       &       1.90  \\
     \hline
\end{longtable}
}
%
%
\end{document}